\documentclass[a4paper, epj]{svjour}

\usepackage[latin1]{inputenc}
\usepackage{epsfig}
\usepackage{graphicx}
\usepackage{units}
\usepackage{amsmath}
\usepackage{amssymb}
\usepackage{cite}

\begin{document}
\title{Friction of the surface plasmon by high-energy particle-hole pairs: 
Are memory effects important?}

\author{Cesar Seoanez\inst{1} \and Guillaume Weick\inst{2,3,4} 
\and Rodolfo A.\ Jalabert\inst{2,3} \and Dietmar Weinmann\inst{2}}
\institute{
  Instituto de Ciencia de Materiales de Madrid, CSIC,
  Cantoblanco E-28049 Madrid, Spain
  \and
  Institut de Physique et Chimie des Matériaux de Strasbourg,
  UMR 7504 (ULP-CNRS), 23 rue du Loess, BP 43, F-67034 Strasbourg Cedex 2,
  France
  \and
  Institut für Physik,
  Universität Augsburg, Universitätsstra{\ss}e 1, D-86135
  Augsburg, Germany
  \and
  Fachbereich Physik, Freie Universität Berlin, Arnimallee 14,
  D-14195 Berlin, Germany
}

\date{\today}

\abstract{
We show that the dynamics of the surface plasmon in metallic 
nanoparticles damped by its interaction with particle-hole excitations can be 
modelled by a single degree of freedom coupled to an environment. In this 
approach, the fast decrease of the dipole matrix elements that couple the 
plasmon to particle-hole pairs with the energy of the excitation 
allows a separation of the Hilbert space into low- and high-energy subspaces
at a characteristic energy that we estimate. A picture of the spectrum
consisting of a collective excitation built from low-energy excitations
which interacts with high-energy particle-hole states can be formalised.
The high-energy excitations yield an approximate description of a
dissipative environment (or ``bath") within a finite confined system.
Estimates for the relevant timescales establish the Markovian character of
the bath dynamics with respect to the surface plasmon evolution for
nanoparticles with a radius larger than about \unit[1]{nm}.
}

\authorrunning{C.\ Seoanez et al.}
\titlerunning{Friction of the surface plasmon by high-energy particle-hole
pairs}
\maketitle

\section{Introduction}
\label{sec_intro}
One of the main questions driving the experimental and theoretical
research of the last twenty years on metallic nanoparticles is 
how large in size do we have to go to observe bulk properties. The
answer of course depends on the physical property under study.
It is furthermore never clearcut. But in many cases the size needed to observe
bulk-like behaviour is surprisingly small. Furthermore, the size dependence of
quantitative features can already be smooth for very small systems. 
For instance, the binding
energy of ${\rm Na}_9$ clusters is quite close to that of ${\rm Na}_8$,
even if these two systems are very different from the molecular point of view
\cite{deHeer}. Such a continuity points towards the relatively
minor importance of the ionic cores and supports the descriptions
based on the jellium model, where the conduction electrons are subject to a
uniform neutralising background \cite{Brack}.
This view has been validated by the evidence of electronic shells
provided by the abundance spectra of alkaline clusters \cite{Knight}.
Within the jellium approximation, the density of electronic states is given by
a bulk-like contribution to which we have to add surface and 
periodic-orbit (or shell) corrections \cite{Brack-book}.

The optical properties of metallic clusters are dominated by the response
of the conduction electrons. Except for the smallest clusters, where
the transitions between single particle levels dictate the optical
response, the optical absorption is dominated by a collective excitation,
the surface plasmon. This resonance is located near the classical Mie frequency
$\omega_{\rm M}=\omega_\textrm{p}/\sqrt{3}$, where $\omega_\textrm{p}=(4\pi
n_{\rm e} e^2/m_{\rm e})^{1/2}$ is the bulk plasma frequency and $e$, 
$m_{\rm e}$, and $n_{\rm e}$ denote the electronic charge, mass, and density, respectively. 
Nevertheless, surface effects lead to a reduction of the surface plasmon frequency with
respect to the bulk value $\omega_{\rm M}$ \cite{Brack, gerchikov, weick_2}. 
The question of how large the size of the nanoparticle has to be
in order to observe a collective excitation in a finite system has
the surprising answer that a very small cluster, like for instance ${\rm Na}_6$, may
already be enough \cite{Selby,Yannouleas-PRA}.

For nanoparticle radii $a$ between about 0.5
and \unit[5]{nm} ($N=8$ to 14000 conduction electrons for the case of Na) the
main effect limiting the lifetime of the resonance is the Landau
damping, i.e., decay into particle-hole pairs. The resulting
linewidth is given by
\begin{equation}
\label{gamma_t}
\gamma_{\rm tot}(a)=\gamma_{\rm i}+\gamma(a)+\gamma^{\rm osc}(a)\,,
\end{equation}
where $\gamma_{\rm i}$ is the intrinsic bulk-like linewidth. The second
term in the right-hand side of \eqref{gamma_t} decreases with the 
na\-no\-par\-ti\-cle size as \cite{kubo, barma, yannouleas}
\begin{equation}
\label{gamma_intro}
\gamma(a)=\frac{3v_{\rm F}}{4a}
g_0\left(\frac{\varepsilon_{\rm F}}{\hbar\omega_{\rm M}}\right)\,,
\end{equation}
with $\varepsilon_{\rm F}=\hbar^2k_{\rm F}^2/2m_{\rm e}$ and $v_{\rm F}$
the Fermi energy and velocity, respectively. The monotonically increasing function
$g_0$ is defined as 
\begin{align}
\label{g_0}
g_0(x)=&\;\frac{1}{12x^2}
\Big\{
\sqrt{x(x+1)} \big(4x(x+1)+3\big)  \nonumber\\
&-3(2x+1) \ln{\left(\sqrt{x}+\sqrt{x+1}\right)} \nonumber\\
&-\Big[ \sqrt{x(x-1)} \big(4x(x-1)+3\big) \nonumber\\
&-3(2x-1) \ln{\left(\sqrt{x}+\sqrt{x-1}\right)}\Big]\Theta(x-1) 
\Big\}\,,  
\end{align}
$\Theta(x)$ being the Heaviside step function.
The last contribution $\gamma^{\rm osc}\sim\cos{(2k_{\rm F}a)}/(k_{\rm F}a)^{5/2}$ 
to the linewidth \eqref{gamma_t} is nonmonotonic in
$a$, and arises from the correlation of the densities of states of the particles
and holes \cite{molina,weick_1}. It becomes relevant for the smallest sizes of the considered
interval, typically for radii in the range 0.5--\unit[1.5]{nm}. 
For radius larger than \unit[5]{nm}
the Landau damping competes with radiation damping in limiting the
lifetime of the resonance. For radii smaller than \unit[0.5]{nm} the
interaction with the ionic background becomes dominant, and the jellium
model no longer provides a useful description.

The use of femtosecond pulsed lasers in pump-probe spectroscopy has rendered
possible to experimentally address the surface plasmon dynamics
\cite{bigot,delfatti,lamprecht,liau}.
The theoretical descriptions built to study this problem
\cite{gerchikov, weick_1, weick_2, EPL} treat the collective coordinate as a special
degree of freedom, which is coupled to an environment constituted by
the degrees of freedom of the relative coordinates. Friction arising
from particle-hole pairs has been studied in bulk systems \cite{Paco}.
In the present work we are interested in the case where the excitation spectrum arises
from a \textit{finite} number of particles. The finite number of electrons makes
the term ``environment" not completely justified in this situation, and
leads us to reformulate the above-mentioned question in dynamical terms
as: How large in size do we need to go to be allowed to describe the relative
coordinates of the electron gas as an environment damping the
collective excitation?

This last question is the main subject of the present paper. In particular,
we determine which are the energies of the electronic excitations that
are active in the damping of the surface plasmon, and which is the characteristic
response time of the large enough electronic environment. The latter is important in
justifying the Markovian approximation that is assumed to describe
the dynamics of the surface plasmon coupled to particle-hole excitations
\cite{sidebands}.

The paper is organised as follows: In Section~\ref{sec_mrpa} we recall the
random phase approximation for the surface plasmon and the
separation of the excitation spectrum in low- and high-energy
particle-hole excitations. In Section~\ref{sec_spslephe}, we show how the
surface plasmon is built from low-energy particle-hole excitations and in
Section~\ref{sec_sraphs} the cutoff energy separating the
above-mentioned low- and high-energy sectors is determined. In Section~\ref{sec_tdrcs}
the time dynamics of the environment giving rise to the damping of the
collective excitation is considered. We end with the conclusions in 
Section~\ref{sec_conclusions}.

\section{Random phase approximation for the surface plasmon}
\label{sec_mrpa}
The understanding of collective excitations in metallic clusters
has benefited from the accumulated knowledge in the related problem
of nuclear physics, i.e., the description of the giant dipolar
resonance \cite{Ring, yanno82}. The simplest many-body approach
yielding collective excitations in a finite system is the Tamm-Dancoff
description, where the excited states are built from the Hartree-Fock
ground state and all possible one particle-one hole (1p-1h) excitations. The
basis considered is then mixed by the matrix elements of the residual
interaction
\begin{equation}
\label{eq:V_res}
V_\textrm{res}=\frac14\sum_{\alpha\beta\gamma\delta}
\bar v_{\alpha\beta\gamma\delta}
c_\alpha^\dagger c_\beta^\dagger c_\delta c_\gamma
-V_\textrm{HF}\,,
\end{equation}
where $c_\alpha^\dagger$ ($c_\alpha$) creates (annihilates) the single-particle state
$|\alpha\rangle$ of the Hartree-Fock problem. In the above equation, 
$\bar v_{\alpha\beta\gamma\delta}=v_{\alpha\beta\gamma\delta}-v_{\alpha\beta\delta\gamma}$ 
with $v_{\alpha\beta\gamma\delta}$ the two-body matrix element of the Coulomb interaction, while
\begin{equation}
V_\textrm{HF}=\frac14\sum_{\alpha\beta}
\bar v_{\alpha\beta\alpha\beta}
c_\alpha^\dagger c_\alpha c_\beta^\dagger c_\beta
\end{equation}
is the Hartree-Fock interaction Hamiltonian.
Diagonalization within this reduced Hilbert space gives the excitation
energies. This task is considerably simplified under the assumption that
the matrix elements ${\bar v}_{\alpha\beta\gamma\delta}$ of the residual 
interaction \eqref{eq:V_res} are separable \cite{Lipparini},
\begin{equation}
\label{separable}
{\bar v}_{\alpha\beta\gamma\delta}=\lambda d_{\alpha\gamma} d_{\delta\beta}^{*}\,,
\end{equation}
where $d_{\alpha\gamma}=\langle \alpha|z|\gamma\rangle$ are dipole matrix elements 
and $\lambda$ is a positive constant characterising the repulsive residual interaction. 
Within the Tamm-Dancoff approximation, the excitation energies are given by the secular equation
\begin{equation}
\label{TDAsum}
\frac{1}{\lambda}=\sum_{ph}\frac{|d_{ph}|^2}{E-\Delta \varepsilon_{ph}}\,,
\end{equation}
where $p$ ($h$) denotes a particle (hole) state above (below) the Fermi level,
and $\Delta \varepsilon_{ph}=\varepsilon_{p}-\varepsilon_{h}$ are 
particle-hole (p-h) excitation energies.

One important drawback of the Tamm-Dancoff approximation is that its
ground state does not include correlations. This is not the case with
the random phase approximation (RPA) where the basis of the
reduced Hilbert space is built from all possible 1p-1h
creations and destructions acting on the ground state. Using again
the separability hypothesis \eqref{separable} for the residual interaction, 
the RPA secular equation 
\begin{equation}
\label{RPAsum_secular}
\frac{1}{\lambda}={\cal S}(E)
\end{equation}
is different from \eqref{TDAsum}. 
Here we have defined the RPA sum
\begin{equation}
\label{RPAsum}
{\cal S}(E)=
\sum_{ph}\frac{2 \Delta\varepsilon_{ph}|d_{ph}|^2}
{E^2-\Delta \varepsilon_{ph}^2}\,.
\end{equation}
The validity of the RPA \cite{Yannouleas-PRA} and of the separability hypothesis
within the time-dependent local density approximation \cite{bapst97} have been
well established, even in small clusters. 

Nuclear physics textbooks (see, e.g., Ref.~\cite{Ring}) show how to solve the 
secular equations \eqref{TDAsum}
and \eqref{RPAsum_secular} graphically under the implicit assumption that
the p-h spectrum is bounded in energy. In that case, the
solutions of the secular equations are merely renormalizations of the
p-h energies $\Delta \varepsilon_{ph}$ except for the
largest excitation energy which corresponds to the collective
mode. However, it is clear that in a metallic cluster the
p-h excitations are not bounded from above on the scale
of the plasma energy. As we will see in the sequel, it is the fast
decay of the dipole matrix elements $d_{ph}$ with the p-h energy that ensures the
applicability of the standard picture.

An ingenious way of describing the collective excitation in
metallic clusters which circumvents the problem of the unbound
p-h spectrum is the separation of the reduced 1p-1h 
RPA Hilbert space in a low-energy sector (the restricted
subspace), containing  p-h excitations with low energy,
and a high-energy sector (the additional subspace)
\cite{yannouleas, yanno82}. The residual interaction gives rise to the
collective excitation as a coherent superposition of a large number
of basis states of the restricted subspace. This excitation energy
lies in the high-energy sector, and the nonvanishing coupling with
p-h states of the additional RPA subspace results in the
broadening of the collective resonance. In this approach the cutoff
energy separating the two subspaces is somehow arbitrary, as long as
it is taken smaller than the resonance energy. The arbitrariness of
the cutoff is not problematic for calculating the lifetime of the
resonance \cite{yannouleas}, but the approximate theoretical description 
of other physical quantities might depend on the cutoff. For example, in 
the approach where the environment is separated from the low-energy 
excitations that compose the collective plasmon excitation, the timescales 
that characterise the dynamics of the electronic environment \cite{sidebands} 
depend on the cutoff energy. Moreover, it was shown in \cite{weick_2} 
that the environment-induced redshift of the resonance frequency 
depends logarithmically on the cutoff. In the following we will provide 
an estimation of the cutoff energy separating the two subspaces.

\section{The surface plasmon as a superposition of low-energy particle-hole
excitations}
\label{sec_spslephe}
The secular equations (\ref{TDAsum}) and (\ref{RPAsum_secular}) depend
crucially on the form of the dipole matrix element $d_{ph}$. Assuming that the p-h states
are confined within a hard-wall sphere, one can decompose \cite{yannouleas}
\begin{equation}
\label{d_ph}
d_{ph} = {\cal A}_{l_p l_h}^{m_p m_h}
{\cal R}_{l_p l_h}(\varepsilon_p , \varepsilon_h)\,.
\end{equation}
The angular part is expressed in terms of Wigner-$3j$ symbols as
\begin{align}
\label{sp_angular}
{\cal A}_{l_p l_h}^{m_p m_h} =
&\,(-1)^{m_p} \sqrt{(2l_p+1)(2l_h+1)}\nonumber\\
&\times
\begin{pmatrix}
l_p & l_h & 1 \\ 
0 & 0 & 0
\end{pmatrix}
\begin{pmatrix}
l_p & l_h & 1 \\ 
-m_p & m_h & 0
\end{pmatrix}
\end{align}
and sets the selection rules $l_p=l_h\pm 1$ and $m_p=m_h$ for
the total and azimuthal angular momenta, respectively. The radial part depends on
the energies of the p-h states as
\begin{equation}
\label{R_ph}
{\cal R}_{l_p l_h}(\varepsilon_p , \varepsilon_h)=\frac{2\hbar^2}{m_{\rm e} a}
\frac{\sqrt{\varepsilon_p \varepsilon_h}}{\Delta\varepsilon_{ph}^2}\,.
\end{equation}

\begin{figure}[t]
\begin{center}
\includegraphics[width=\columnwidth]{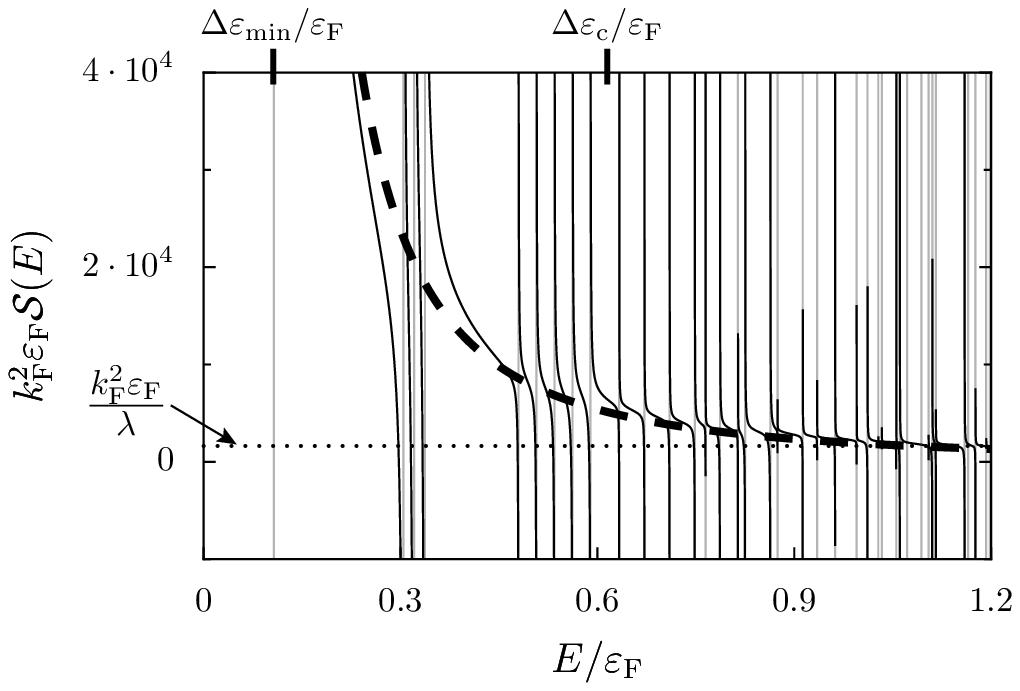}
\caption{\label{sum}%
RPA sum \eqref{RPAsum_inter} for a ${\rm Na}$ nanoparticle with $k_{\rm
F}a=30$ (solid black line). The p-h excitation energies $\Delta\varepsilon_{ph}$
(represented by vertical grey lines) have been
obtained from the semiclassical spectrum \eqref{Ediff}.
There are about 15000 degenerate excitations in the interval shown, and the smallest excitation
$\Delta\varepsilon_\textrm{min}\simeq\varepsilon_\textrm{F}/10$ has a degeneracy
${\cal N}=380$. 
The dashed line takes only into account the contribution of 
$\Delta\varepsilon_\textrm{min}$ (see Eq.~\eqref{RPA_mini}). The cutoff energy
$\Delta\varepsilon_\textrm{c}$ separating the two RPA subspaces is also shown
in the figure (see Eq.~\eqref{estimatedlec3}). Above
$\Delta\varepsilon_\textrm{c}$ the two sums are close to each other, showing
that for high energies the RPA sum is essentially given by the contributions
coming from the low-energy p-h excitations. The horizontal dotted line indicates
the position of the coupling constant $\lambda$ entering the RPA secular
equation \eqref{RPAsum_secular}, according to the estimate
\eqref{lambda}.}
\end{center}
\end{figure}

With the help of equations \eqref{d_ph}--\eqref{R_ph} and using the appropriate
dipole selection rules, we can write the RPA sum \eqref{RPAsum} as
\begin{equation}
\label{RPAsum_inter}
{\cal S}(E)=
\frac{32k_{\rm F}^{-2}}{3}\left(\frac{\varepsilon_{\rm F}}{k_{\rm F}a}\right)^2
\sum_{\substack{n_h, l_h\\n_p, l_p=l_h\pm1}}
f_{l_p}\frac{\varepsilon_p\varepsilon_h}
{(E^2-\Delta\varepsilon_{ph}^2)\Delta\varepsilon_{ph}^3}\,,
\end{equation}
where $n_h$ and $n_p$ are radial quantum numbers. We have defined
$f_{l_p}=l_h+1$ if $l_p=l_h+1$ and $f_{l_p}=l_h$ if $l_p=l_h-1$. Using the
semiclassical quantisations \eqref{qcondition2} and \eqref{Ediff} of
Appendix~\ref{app_Emin}, we finally
obtain the result shown in Figure~\ref{sum}.

In the figure we show (solid black line) the resulting $E$-dependence of 
\eqref{RPAsum_inter}. The crossings of this curve with the horizontal dotted one with 
height $1/\lambda$ yield the excitation spectrum within the RPA. The lowest p-h
energy estimated in Appendix~\ref{app_Emin} is 
\begin{equation}
\label{SpacingGuillaume}
\Delta\varepsilon_\textrm{min}\simeq\frac{\varepsilon_{\rm F}}{k_{\rm
F}a/\pi}\,.
\end{equation}
Whenever the energy $E$ coincides with a p-h
excitation energy $\Delta\varepsilon_{ph}$, we have a divergence in 
${\cal S}(E)$ (see vertical grey lines in Fig.~\ref{sum}). For the lowest energies $E$ of the
interval considered the sum is dominated by the term associated with
the divergence closest to $E$. On the other hand, for the
largest energies $E$, the fast decay of $d_{ph}$ 
with $\Delta\varepsilon_{ph}$ means that the divergences are only
relevant for energies extremely close to them. Away from the
divergences the sum is dominated by the contributions arising from
the low-energy p-h excitations. 

To gain physical insight and proceed further in the analysis of the 
electron dynamics in metallic nanoparticles, we introduce
the typical dipole matrix element for states separated by a given energy
difference $\Delta\varepsilon$,
\begin{align}
\label{Dmieq}
d^\textrm{p-h}(\Delta\varepsilon)&=
\Bigg[\frac{1}{\rho^\textrm{p-h}(\Delta\varepsilon)}
\sum_{ph}{|d_{ph}|}^2\delta(\Delta\varepsilon-\Delta\varepsilon_{ph})\Bigg]^{1/2}\nonumber\\
&=
\Bigg[\frac{1}{\rho^\textrm{p-h}(\Delta\varepsilon)}
\int_{\varepsilon_{\rm F}-\Delta\varepsilon}^{\varepsilon_{\rm F}} {\rm
d}\varepsilon\,
{\cal C}(\varepsilon,\Delta\varepsilon)\Bigg]^{1/2}\,. 
\end{align}
Here, we have introduced the local density of dipole matrix elements
\begin{equation}
\label{lddme}
{\cal C}(\varepsilon,\Delta\varepsilon)=
\sum_{ph}|d_{ph}|^2\delta(\varepsilon-\varepsilon_h)
\delta(\varepsilon+\Delta\varepsilon-\varepsilon_p)\,.
\end{equation}
As shown in Appendix~\ref{app_A}, this can be expressed as 
\begin{equation}
\label{C}
{\cal C}(\varepsilon,\Delta\varepsilon)=
\frac{1}{3\pi^2}\frac{a^2}{\Delta\varepsilon^2}
F\left(\frac{\varepsilon}{\Delta\varepsilon}\right)\,,
\end{equation}
where 
\begin{equation}
\label{F}
F(x)=(2x+1)\sqrt{x(x+1)}-\ln{\left(\sqrt{x}+\sqrt{x+1}\right)}\,.
\end{equation}
For $\Delta\varepsilon\ll\varepsilon$, equation \eqref{C} simplifies to 
\begin{equation}
\label{Csmallde}
{\cal C}(\varepsilon,\Delta\varepsilon)\simeq\frac{2a^2}{3\pi^2}
\frac{\varepsilon^2}{\Delta\varepsilon^4}\,.
\end{equation}

In expression \eqref{Dmieq}, 
\begin{equation}
\label{rph1}
\rho^\textrm{p-h}(\Delta\varepsilon)=\sum_{ph}
\delta(\Delta\varepsilon_{ph}-\Delta\varepsilon) 
\delta_{l_h,l_p\pm1}\delta_{m_h,m_p}
\end{equation}
is the density of p-h excitations with
energy $\Delta\varepsilon$ respecting the dipole selection rules.
In order to simplify the presentation, we do not consider
spin degeneracy factors. In Appendix~\ref{app_B} we show that for $\Delta\varepsilon\ll
\varepsilon_\textrm{F}$ we have
\begin{equation}
\label{rph3}
\rho^\textrm{p-h}(\Delta\varepsilon)\simeq\frac{\left(k_{\rm F}a\right)^4}{4\pi^2}
\frac{\Delta\varepsilon}{\varepsilon_{\rm F}^2}\,,
\end{equation}
and therefore in such a limit the typical matrix element \eqref{Dmieq} can be
approximated by
\begin{equation}
\label{D1}
d^\textrm{p-h}(\Delta\varepsilon)\simeq
\frac{2\sqrt{2}k_{\rm F}^{-1}}{\sqrt{3}k_{\rm F}a}
\left(\frac{\varepsilon_{\rm F}}{\Delta\varepsilon}\right)^2\,.
\end{equation}
As a check for our estimation of the typical dipole matrix element, we have
evaluated the energy-weighted sum rule
$\sum_{ph}\Delta\varepsilon_{ph}|d_{ph}|^2$ using \eqref{D1} and obtained about 70\% of the exact
result $(3/4\pi)\hbar^2N/2m_{\rm e}$ \cite{Brack}. This is quite reasonable regarding all the
approximations we made to obtain \eqref{D1}.

In order to illustrate the importance of the low-energy p-h excitations, 
we present in Figure~\ref{sum} (dashed line) the
contribution to the RPA sum coming only from
the infrared p-h excitation energy with the
appropriate degeneracy factor $\cal N$ (see Eq.~\eqref{deg1} in App.~\ref{app_Emin}). 
Indeed, we can estimate \eqref{RPAsum} as 
\begin{equation}
{\cal S}(E)\approx{\cal N}\times\frac{2\Delta\varepsilon_{\rm min}
\left[d^\textrm{p-h}(\Delta\varepsilon_{\rm min})\right]^2}
{E^2-\Delta\varepsilon_{\rm min}^2}\,.
\end{equation}
With the results \eqref{SpacingGuillaume}, \eqref{D1}, and \eqref{deg1}, we
obtain
\begin{equation}
\label{RPA_mini}
{\cal S}(E)\approx\frac{64k_{\rm F}^{-2}}{3\pi^5}(k_{\rm
F}a)^3\frac{\varepsilon_{\rm F}}{E^2-(\varepsilon_{\rm F}\pi/k_{\rm
F}a)^2}\,.
\end{equation}
While in the lower 
part of the energy interval in Figure~\ref{sum} the two curves exhibit considerable 
discrepancies, in the second half of the interval they are very close (except of course at the
divergences). Since the collective excitation is found in this last interval, we see that it
is mainly the low-energy p-h excitations that are relevant
for the definition of the collective excitation. 

Since the resonance energy is known from experiments, we can obtain
the value of the coupling constant $\lambda$. 
Indeed, using the estimate \eqref{RPA_mini} evaluated at $E=\hbar\omega_{\rm
M}$ in the secular equation \eqref{RPAsum_secular}, we obtain
\begin{equation}
\label{lambda}
\frac{1}{\lambda}\simeq
\frac{64k_{\rm F}^{-2}}{3\pi^5}(k_{\rm F}a)^3
\frac{\varepsilon_{\rm F}}{(\hbar\omega_{\rm M})^2}
\end{equation}
to leading order in $k_{\rm F}a\gg1$. Notice that this result is consistent with
the estimate obtained from the energy-weighted sum rule given in \cite{yannouleas}.
For the case studied in Figure \ref{sum}, our estimate \eqref{lambda} yields 
$k_{\rm F}^2\varepsilon_{\rm F}/\lambda\approx1600$.
As the radius $a$ increases the lowest p-h energy decreases like $1/a$, the
degeneracy at this value increases as $a^2$, and the density of
p-h excitations contributing to 
(\ref{RPAsum}), $\rho^\textrm{p-h}(\Delta\varepsilon)$, grows as $a^4$. This
increase in the number of excitations contributing to the sum is
partially cancelled by the $1/a$ behaviour of the typical dipole matrix
element, resulting in a decrease of the coupling constant $\lambda$
proportional to the number of particles in the cluster, $\lambda\sim
1/a^3$, and a value of the plasmon frequency which remains almost
unaffected. Therefore, for a larger nanoparticle size, the divergences 
shown in Figure~\ref{sum} would be more dense in energy,
starting at a lower energy, and the vertical scale would increase as
$a^3$, thus obtaining a similar value for the plasmon excitation
energy $\hbar\omega_{\rm M}$.

\section{Separation of the reduced and additional particle-hole subspaces}
\label{sec_sraphs} 
As we have shown in the last section, the high-energy
part of the p-h spectrum is not crucial for the
determination of the energy of the collective excitation. In what
follows we make this statement more quantitative and estimate the
upper-bound cutoff $\Delta\varepsilon_\textrm{c}$ of the low-energy excitations that we
need in order to obtain a stable position of the surface plasmon. 

In order to obtain a quantitative estimate of the cutoff, we require that by changing it
from $\Delta\varepsilon_\textrm{c}$ to $(3/2)\Delta\varepsilon_\textrm{c}$, 
the position of the plasmon changes only by a fraction of its linewidth $\gamma$, 
the smallest energy scale with experimental significance. Our criterion leads to
the condition
\begin{align}
\label{estimatedlec2}
{\cal S}_{\Delta\varepsilon_{\rm c}}(\hbar\omega_{\rm M})=
{\cal S}_{\frac32\Delta\varepsilon_{\rm c}}(\hbar\omega_{\rm M}+\hbar\gamma)\,,
\end{align}
with the RPA sum $\cal S$ that has been defined in \eqref{RPAsum}. The
additional subscript refers to the upper bound of the p-h energies.

The left-hand side of \eqref{estimatedlec2} can be estimated according to 
\begin{equation}
{\cal S}_{\Delta\varepsilon_{\rm c}}(\hbar\omega_{\rm M})
\simeq\int_{\Delta\varepsilon_\textrm{min}}^{\Delta\varepsilon_\textrm{c}}
{\rm d}\Delta\varepsilon\,
\rho^\textrm{p-h}(\Delta\varepsilon)\frac{2\Delta\varepsilon
\left[d^\textrm{p-h}(\Delta\varepsilon)\right]^2}{(\hbar\omega_{\rm
M})^2-\Delta\varepsilon^2}\,,
\end{equation}
with $d^\textrm{p-h}$ and $\rho^\textrm{p-h}$ as defined in \eqref{Dmieq} 
and \eqref{rph1}, respectively. Using \eqref{rph3} and \eqref{D1}, we obtain 
\begin{equation}
{\cal S}_{\Delta\varepsilon_{\rm c}}(\hbar\omega_{\rm M})\simeq\frac{4k_{\rm F}^{-2}}{3\pi^2}
\left(k_{\rm F}a\frac{\varepsilon_{\rm F}}{\hbar\omega_{\rm M}}\right)^2
\left(\frac{1}{\Delta\varepsilon_{\rm min}}
-\frac{1}{\Delta\varepsilon_{\rm c}}
\right)
\end{equation}
to leading order in $k_{\rm F}a$.
Similarly, 
\begin{align}
{\cal S}_{\frac32\Delta\varepsilon_{\rm c}}(\hbar\omega_{\rm M}+\hbar\gamma)
\simeq&\;\frac{4k_{\rm F}^{-2}}{3\pi^2}
\left(k_{\rm F}a\frac{\varepsilon_{\rm F}}{\hbar\omega_{\rm M}}\right)^2
\left(1-\frac{2\gamma}{\omega_{\rm M}}\right)\nonumber\\
&\times\left(\frac{1}{\Delta\varepsilon_{\rm min}}
-\frac{2}{3\Delta\varepsilon_{\rm c}}\right)\,.
\end{align}
Using the expressions \eqref{gamma_intro} and
\eqref{SpacingGuillaume} for $\gamma$ and $\Delta\varepsilon_{\rm min}$, respectively, 
finally yields according to the criterion \eqref{estimatedlec2} the cutoff energy 
\begin{equation}
\label{estimatedlec3} 
\Delta\varepsilon_\textrm{c}\simeq\frac{\pi}{9g_0(\varepsilon_{\rm F}/\hbar\omega_{\rm M})}
\hbar\omega_\textrm{M}\,,
\end{equation}
with the function $g_0$ defined in \eqref{g_0}.
For Na clusters, we have $\varepsilon_\textrm{F}/\hbar\omega_\textrm{M}=0.93$ and
our criterion yields a value 
$\Delta\varepsilon_{\rm c} \simeq (3/5)\varepsilon_{\rm F}$.

We have verified the robustness of our criterion \eqref{estimatedlec2} by
exploring different physical parameters, like the size of the cluster. 
Indeed, it can be seen in Figure~\ref{sum} that above
$\Delta\varepsilon_\textrm{c}$ the RPA sum evaluated from \eqref{RPAsum_inter}
(solid line) and the estimate \eqref{RPA_mini} (dashed line) are close to each other, showing
that for high energies the RPA sum is essentially given by the contributions
coming from the low-energy p-h excitations. We have checked that this feature of 
$\Delta\varepsilon_{\rm c}$ is independent of the size $a$ of the nanoparticle.

In order to have a well-defined collective excitation, the cutoff
energy $\Delta\varepsilon_{\rm c}$ must obviously be larger than the minimal p-h
excitation energy \eqref{SpacingGuillaume}. 
For Na, we find that this condition is already verified with only $N=20$ conduction
electrons in the nanoparticle, in agreement with the experiments of reference 
\cite{Selby} and the numerical calculations of reference \cite{Yannouleas-PRA}.

The splitting in low and high-energy p-h excitations is
important in order to justify the separation of the electronic degrees
of freedom into centre-of-mass and relative coordinates
\cite{gerchikov, weick_1, weick_2, EPL}. Within such an approach, the Hamiltonian of
the electronic system is written as 
\begin{equation}
\label{H_decompo}
H = H_{\rm cm}+H_{\rm rel}+H_{\rm c}\,. 
\end{equation}
The first term describes the centre-of-mass motion as a harmonic
oscillator. The second term describes the relative coordinates as
independent fermions in the effective mean-field potential. The
coupling term $H_{\rm c}$ describes the creation or annihilation of
a surface plasmon by destruction or creation of p-h pairs. Its strength is given
by the coupling constant $\Lambda=(\hbar m_{\rm e}\omega_{\rm M}^3/2N)^{1/2}$ and the 
dipole matrix element $d_{ph}$ \cite{weick_2}.
It is important to remark that, even if we have used the separability
hypothesis of the residual interaction to justify the decomposition
\eqref{H_decompo}, the model that is put forward does not rely on such an
approximation.

\section{Dynamics of the relative-coordinate system}
\label{sec_tdrcs}
The decomposition \eqref{H_decompo} of $H$ 
suggests to treat the collective coordinate as a simple system of
one degree of freedom which is coupled to an environment with many
degrees of freedom. The latter are the relative coordinates
described by $H_{\rm rel}$. This is the approach taken in
\cite{sidebands}. In this picture, the time evolution of the
centre-of-mass system (i.e., the surface plasmon) strongly depends on the dynamics of the
relative-coordinate system. Such a dynamics is characterised by a 
correlation function which can be written at zero temperature as\cite{sidebands}
\begin{equation}
\label{Cenv1}
C(t)=\Lambda^2\sum_{ph}
|d_{ph}|^2 {\rm e}^{{\rm i}\Delta\varepsilon_{ph}t/\hbar}
\Theta(\Delta\varepsilon_{ph}-\Delta\varepsilon_\textrm{c})\,.
\end{equation}
In order to evaluate the above expression, it is helpful to introduce its
Fourier transform
\begin{equation}
\label{Sigma}
\Sigma(\Delta\varepsilon)=\frac{2\pi}{\hbar}\Lambda^2
\sum_{ph}|d_{ph}|^2\delta(\Delta\varepsilon-\Delta\varepsilon_{ph})
\Theta(\Delta\varepsilon_{ph}-\Delta\varepsilon_\textrm{c})
\end{equation}
which has been calculated in \cite{weick_2}. 
Finite temperatures were shown to result in a small quadratic correction.
Consistently with the results of the preceding sections, we employ our low
energy estimates \eqref{rph3} and \eqref{D1} to obtain
\begin{equation}
\label{Sigma_result}
\Sigma(\Delta\varepsilon)\simeq\frac{3v_{\rm F}}{4a}
\left(\frac{\hbar\omega_{\rm M}}{\Delta\varepsilon}\right)^3
\Theta(\Delta\varepsilon-\Delta\varepsilon_\textrm{c})\,.
\end{equation}
This result is consistent with the one of reference \cite{weick_2} 
(see Eq.~(34) in there) in the limit 
$\Delta\varepsilon\ll\hbar\omega_{\rm M}$ and for zero temperature.

In principle we could calculate $C(t)$ by taking the inverse Fou\-ri\-er
transform of \eqref{Sigma_result}. The
decay of $C(t)$ for very long times is dominated by the
discontinuity of $\Sigma$ at $\Delta\varepsilon_\textrm{c}$. This is somehow
problematic since $\Delta\varepsilon_\textrm{c}$ can only be estimated as we did in
Section~\ref{sec_sraphs}, and since the functional form of the long
time decay depends on how sharply the cutoff is implemented.
However, it is important to realise that it is not the very long
time behaviour that determines the relevant decay of the correlation function, but rather 
the typical values at which $C(t)$ is reduced by an important factor
from its initial value $C(0)$. We then estimate the correlation
time as the mean decay time of $C(t)$, 
\begin{align}
\label{estimatortau}
\langle\tau_{\rm cor}\rangle&=\left|\int_{0}^{\infty}{\rm d}t\,t\frac{\rm d}{{\rm d}t}
   \left(\frac{C(t)}{C(0)}\right)\right|\nonumber\\
   &=\frac{1}{C(0)} \left| \int_{0}^{\infty}{\rm d}t\, C(t) \right|\,.
\end{align}
If $C(t)$ were an exponentially decreasing function, $\langle\tau_{\rm
cor}\rangle$ would simply reduce to the inverse of the decay rate. 
Using the definition \eqref{Sigma}, we get
\begin{equation}
\label{estimatortau2}
\langle\tau_\textrm{cor}\rangle=
\frac{\hbar\displaystyle
\int_{\Delta\varepsilon_\textrm{c}}^{\infty}\textrm{d}\Delta\varepsilon
\Sigma(\Delta\varepsilon)/\Delta\varepsilon}
{\displaystyle
\int_{\Delta\varepsilon_\textrm{c}}^{\infty}\textrm{d}\Delta\varepsilon
\Sigma(\Delta\varepsilon)}\,.
\end{equation}
Given the fast decay of the function $\Sigma\sim1/\Delta\varepsilon^3$, the
above integrals are dominated by their lower limit $\Delta\varepsilon_\textrm{c}$ and we have
\begin{equation}
\label{estimatortau3}
\langle\tau_{\rm cor}\rangle\simeq\frac{2}{3}\frac{\hbar}{\Delta\varepsilon_\textrm{c}}\,.
\end{equation}
Since $\Delta\varepsilon_\textrm{c}$ is of the order of 
$(3/5)\varepsilon_\textrm{F}$, we see that the
response time (or correlation time) of the electronic environment is of the order of its
inverse Fermi energy.

The estimation of the characteristic response time of the
electronic environment is crucial in justifying the Mar\-ko\-vian
approximation used in \cite{sidebands}. In that work, the
degrees of freedom corresponding to the relative coordinates were
integrated out and treated as an incoherent heat bath that acts on
the collective coordinate. Such an approach relies on the fast
response of the environment as compared to the time evolution of the
surface plasmon. The typical scale for the latter is the inverse of the decay
rate, $\tau_{\rm sp}=1/\gamma$. Using \eqref{gamma_intro} and \eqref{estimatedlec3}, 
we therefore have
\begin{equation}
\label{tratio1}
\frac{\tau_{\rm sp}}{\langle\tau_{\rm cor}\rangle}=
\frac{\pi}{9[g_0(\varepsilon_{\rm F}/\hbar\omega_{\rm M})]^2}
\frac{\hbar\omega_{\rm M}}{\varepsilon_{\rm F}}k_\textrm{F} a\,.
\end{equation}
For the example of Na nanoparticles worked in Section~\ref{sec_spslephe}, we have
$\tau_\textrm{sp}/\langle\tau_{\rm cor}\rangle\simeq k_{\rm F}a$. 
This is a safe limit since in not too small nanoparticles, $k_{\rm F}a\gg1$. 
As the size of the cluster increases, the
applicability of the Markovian approximation is more justified. This is
expected since the electronic bath has more and more degrees of
freedom, approaching an ``environment" in the sense of quantum dissipation.
Since the physical parameters of alkaline nanoparticles entering \eqref{tratio1} 
are close to that of noble-metal clusters, the dynamics of the surface plasmon
can be expected to be Markovian in that case too.

\section{Conclusions}
\label{sec_conclusions}
We have studied the role of particle-hole excitations
on the dynamics of the surface plasmon. A key concept in
this analysis is the separation into low-energy excitations which lead
to the collective excitation once they are mixed by the residual
interaction, and high-energy excitations that act as an environment
damping the resonance. Using the random phase approximation and
assuming the separability of the residual interaction, we have
established a criterion for estimating the cutoff energy separating
the low- and high-energy subspaces. The resulting cutoff energy is
approximately $(3/5)\varepsilon_{\rm F}$ for the case of Na nanoparticles.

Since the number of electrons in the cluster is finite, the
assumption that the high-energy particle-hole excitations act on the
collective excitation as an environment, introducing friction in its dynamics, 
may be questionable. What
settles this issue is the ratio between the typical evolution time 
of the collective excitation and the one of the high-energy
particle-hole excitations. The former is given by the inverse
of the plasmon linewidth, while the latter is obtained from the
decay of the correlation function of the environment. We have found that
this ratio improves with increasing cluster size. Even for a
small cluster with $a=\unit[1]{nm}$, the ratio is
approximately 10, justifying the use of the Markovian
approximation which assumes a fast time evolution of the environment with
respect to the one of the collective excitation. 

The relevance of memory effects in the electronic dynamics of small
clusters is of current interest, due to the advance in time-resolved
experimental techniques \cite{bigot,delfatti,lamprecht,liau}. 
First-principle calculations have recently addressed this issue by
comparing time-dependent density functional theories with and
without memory effects for small gold clusters \cite{KB06}. For very
small clusters ($N<8$) memory effects were shown to be
important. We stress that the memory considered in \cite{KB06}
is that of the electron gas as a whole, while we are concerned in
this work with the memory arising from the dynamical evolution of
the relative-coordinate subsystem. It would be interesting to
consider cluster sizes intermediate between the ones considered in
the present work and those of reference \cite{KB06} in order to study the
emergence of memory effects.

\begin{acknowledgement} 
We thank F.\ Guinea, G.-L.\ Ingold, and E.\ Mariani for helpful discussions.
We acknowledge financial support from
the ANR,
the Deutsche Forschungsgemeinschaft,
the EU through the MCRTN program,
the French-German PAI program Procope, 
and the Ministerio de Educaci\'on y Ciencia (MEC).
\end{acknowledgement}

\appendix
\section{Lowest energy of the particle-hole spectrum}
\label{app_Emin}
If we consider the cluster as a hard-wall sphere of radius $a$, its
eigenstates are given in terms of spherical Bessel functions. Using
the large $ka$ expansion of the latter (semiclassical high-energy limit), 
the quantisation condition reads
\begin{equation}
\label{qcondition}
ka=\pi\left(\frac{l}{2}+n\right)
\end{equation}
with $l$ and $n$ non-negative integers.
The energy of a single-particle (hole) state is related to its wavevector
$k_{p(h)}$, its total angular momentum $l_{p(h)}$, and its radial quantum number
$n_{p(h)}$ as
\begin{align}
\label{qcondition2}
\varepsilon_{p(h)}&=\frac{\hbar^2k_{p(h)}^2}{2m_{\rm e}}\nonumber\\
&=\varepsilon_\textrm{F}\left(\frac{\pi}{k_{\rm F}a}\right)^2
\left(\frac{l_{p(h)}}{2}+n_{p(h)}\right)^2\,.
\end{align}
Thus, the energy of a p-h excitation entering the RPA sum \eqref{RPAsum_inter} is
\begin{align}
\label{Ediff}
\Delta\varepsilon_{ph}=
\varepsilon_\textrm{F}\left(\frac{\pi}{k_{\rm F}a}\right)^2
&\left(\frac{l_p-l_h}{2}+n_p-n_h\right)\nonumber\\
\times&\left(\frac{l_p+l_h}{2}+n_p+n_h\right)\,.
\end{align}

Notice that using the exact quantum mechanical spectrum in \eqref{RPAsum_inter}
would not change significantly the result depicted in Figure \ref{sum}, since
the approximation \eqref{qcondition} is very reliable for states close to the
Fermi energy. We have also checked that
generating the p-h excitation energies randomly in the RPA sum \eqref{RPAsum}
does not affect the physical picture of Figure \ref{sum}. Indeed, the main
ingredient to understand such a picture is the fast decay of the dipole matrix
element with the p-h energy.

The expressions for the dipole matrix element \eqref{R_ph} as well as for the
typical dipole matrix element \eqref{D1} diverge in the limit of a small p-h
energy. It is therefore crucial for our analysis to determine the appropriate 
minimal p-h energy $\Delta\varepsilon_{\rm min}$ that renders this divergence unphysical.

This can be achieved by imposing the dipole selection rules in \eqref{Ediff} and
that the energy difference is minimal. The first condition dictates
that $l_h=l_p±1$ and $m_h=m_p$. Therefore there are two ways of
obtaining the minimal energy difference: $n_p=n_h$ with $l_p=l_h+1$
and $n_p=n_h+1$ with $l_p=l_h-1$. In both cases we have
\begin{equation}
\label{dleminGuillaume}
\Delta\varepsilon_\textrm{min}\simeq
\frac{\varepsilon_{\rm F}}{k_{\rm F}a/\pi}\frac{k_h}{k_{\rm F}}\,.
\end{equation}
Since we are interested in states close to the Fermi level, we
can simplify \eqref{dleminGuillaume} to expression \eqref{SpacingGuillaume}.

If we consider sodium clusters with $k_\textrm{F}a=30$ ($a=\unit[3.3]{nm}$ and
$N\simeq4000$ conduction electrons per spin direction), we have 
$\Delta\varepsilon_\textrm{min}\approx\varepsilon_{\rm F}/10$. 
This is a much larger energy than the lowest one 
we can observe in the numerically generated excitation spectrum
(see Fig.~1 in Ref.~\cite{molina}). However, the two results are
reconciled once we take into account the large degeneracy yielded
by our approximate quantisation condition (\ref{qcondition}).

The degeneracy of p-h excitations with minimal energy is
given by twice the number of pairs ($l_h,n_h$) compatible with
$k_h=k_{\rm F}$ and $\Delta\varepsilon_{ph}=\Delta\varepsilon_{\rm min}$.
Indeed, we have seen that there are two
possible particle states $p$ starting from $h$ and verifying the
above-mentioned conditions. For each $n$ between 1 and $k_{\rm F}a/\pi$,
there is a value of $l=2(k_{\rm F}a/\pi-n)$ and therefore the number of
degenerate p-h excitations with energy $\Delta\varepsilon_{\rm min}$ is
\begin{align}
\label{deg1}
{\cal N}&=2\sum_{n=0}^{k_{\rm F}a/\pi}(2l+1)\nonumber\\
&=2\sum_{n=0}^{k_{\rm F}a/\pi}\left[4\left(\frac{k_{\rm F}a}{\pi}-n\right)+1\right]
\nonumber\\
&\simeq 4\left(\frac{k_{\rm F}a}{\pi}\right)^2\,.
\end{align}
This degeneracy factor has to be included in Figure \ref{sum}, and it
is crucial for the determination of the collective excitation.

\section{Local density of the dipole matrix element}
\label{app_A}
Equation \eqref{lddme} defines the local density of dipole matrix
elements connecting states at energies $\varepsilon$ and $\varepsilon+\Delta\varepsilon$.
We used particle and hole states in our definition, since this is the main
interest of our work. But note that the calculation presented in this appendix is
not restricted to that case and can be easily extended to any states. 
The result would be of course unchanged.

Local densities of matrix elements of arbitrary operators have been thoroughly studied as they
can be easily connected with physical properties, ranging from
far-infrared absorption in small particles \cite{MR97} to electronic
lifetimes of quantum dots \cite{GJS04}.
A semiclassical theory for the local density of matrix elements has been
developed \cite{FP86,W87,EFMW92}, where \eqref{lddme} can be
expressed as a smooth part given by correlations along classical trajectories
plus a periodic orbit expansion. We will not follow here this general
procedure, but use the simple form of the dipole matrix elements \eqref{d_ph} for
states confined in a hard-wall sphere 
and the semiclassical approximation applied to the radial (fixed $l$)
problem \cite{molina,weick_1}. 

Introducing the $l$-fixed density of states, which in leading order
in $\hbar$ is given by
\begin{equation}
\label{semiclassical_DOS}
\varrho_{l}(\varepsilon)=\frac{\sqrt{2m_{\rm
e}a^2\varepsilon/\hbar^2-(l+1/2)^2}}{2\pi\varepsilon}\,,
\end{equation}
we can write with the help of equations \eqref{d_ph}--\eqref{R_ph}
\begin{align}
\label{CofEA1}
  {\cal C}(\varepsilon,\Delta\varepsilon)=&
  \left(\frac{2\hbar^2}{m_{\rm
  e}a}\right)^2\frac{\varepsilon\varepsilon'}{3\Delta\varepsilon^4}\nonumber\\
  &\times\sum_{l_h=0}^{l_{\rm
  max}}\varrho_{l_h}(\varepsilon)\left[(l_h+1)
  \varrho_{l_h+1}(\varepsilon')+l_h\varrho_{l_h-1}(\varepsilon')\right]
\end{align}
where $l_{\rm max}$ is the maximum allowed $l_h$ for an energy $\varepsilon$, while 
$\varepsilon'=\varepsilon+\Delta\varepsilon$. In the
semiclassical limit we can take $l_h\simeq l_h+1\gg1$ and convert the
sum into an integral. Thus, 
\begin{align}
  {\cal C}(\varepsilon,\Delta\varepsilon) &\simeq
   \frac{1}{6\pi^2}\left(\frac{2\hbar^2}{m_{\rm e}
   a}\right)^2\frac{1}{\Delta\varepsilon^4}\nonumber\\
   &\times\int_{0}^{\sqrt{\frac{2m_{\rm e} a^2}{\hbar^2}\varepsilon}}{\rm
   d}l\,l\sqrt{\frac{2ma^2}{\hbar^2}\varepsilon-l^2}
   \sqrt{\frac{2ma^2}{\hbar^2}\varepsilon'-l^2}\,.
\end{align}
Performing the remaining integral over the angular momentum $l$ finally yields
the result \eqref{C}.

\section{Density of particle-hole excitations}
\label{app_B}
The density of p-h excitations with energy $\Delta\varepsilon$ is defined
in \eqref{rph1} and can be written as
\begin{align}
\label{rph2}
\rho^\textrm{p-h}(\Delta\varepsilon)=&
\int_{\varepsilon_\textrm{F}-\Delta\varepsilon}^{\varepsilon_\textrm{F}}
{\rm d}\varepsilon_h\sum_{l_h}(2l_h+1)\varrho_{l_h}(\varepsilon_h)\nonumber\\
&\times
\left[\varrho_{l_h+1}(\varepsilon_h+\Delta\varepsilon)
+\varrho_{l_h-1}(\varepsilon_h+\Delta\varepsilon)\right]\,.
\end{align}
Using the semiclassical density of states \eqref{semiclassical_DOS}
and performing the sum in the limit $l_h\gg1$, we obtain
\begin{equation}
\rho^\textrm{p-h}(\Delta\varepsilon)\simeq
\frac{\Delta\varepsilon^2}{8\pi^2}\left(\frac{2m_{\rm e}a^2}{\hbar^2}\right)^2
\int_{\varepsilon_\textrm{F}-\Delta\varepsilon}^{\varepsilon_\textrm{F}}
{\rm d}\varepsilon_h\frac{F(\varepsilon_h/\Delta\varepsilon)}
{\varepsilon_h (\varepsilon_h+\Delta\varepsilon)}\,,
\end{equation}
where the function $F$ has been defined in \eqref{F}. Performing the remaining
integral over the hole energy in the limit $\Delta\varepsilon\ll\varepsilon_{\rm
F}$ is straightforward and leads to the result \eqref{rph3}.


\end{document}